\documentclass{elsart}

\usepackage{epsfig}

\begin{document}

\begin{frontmatter}



\title{Breaking the scale invariance of the primordial spectrum or not: the new WMAP data}


\author{Davor Palle \thanksref{1}}
\thanks[1]{Fulbright research scholar on leave from Rugjer Bo\v skovi\' c
Institute, Zagreb, Croatia}
\address{Astronomy Department, New Mexico State University, Box 30001 \\
 Department 4500, Las Cruces, NM 88003-0001}

\begin{abstract}
It seems that new WMAP data require a fit with a 
primordial spectrum containing small negative tilt index
in addition to the featureless Harrison-Zeljdovi\v c-Peebles
spectrum thus implying a broken scale invariance.
We show that the data could be otherwise interpreted by 
a scale invariant primordial spectrum with
the scale noninvariant evolution of density contrast using 
the Press-Schechter formalism. The estimate of
the acceleration parameter, as a source of the 
inhomogeneity of spacetime, is made
by searching for the minima of the deviation measure defined
by the Press-Schechter mass functions for this interpretation
compared to the assumptions implicit in the WMAP fit.
\end{abstract}

\end{frontmatter}


There is a wide agreement in cosmology that the $\Lambda CDM$ model and 
the inflationary scenario describe in detail a variety of
cosmological data. However, general relativity and inflation alone cannot
resolve and determine the cosmological constant. The dynamics of
the hypothetical inflaton scalar at the scale of $O(10^{-31} cm)$
defines the form but not the magnitude of
the primordial fluctuation spectrum at the scale
of photon background decoupling $O(10^{25} cm)$.

The fit of three years of WMAP data \cite{r1} results in low 
mass denity $\Omega_{m}=0.266$ and flat geometry $\Omega_{\Lambda}=
1-\Omega_{m}$, but the large scale primordial fluctuation remains
disasterrously small. This is the confirmation of the COBE and
WMAP one year data and there is little chance that future measurements
of WMAP or Planck will alter the observed extremely low large scale power of
the fluctuations. The negative contribution of the integrated Sachs-Wolfe effect
to $TT$ spectrum, as a possible resolution of the problem, presumes a
negative cosmological constant, in apparent contradiction with the overall
WMAP fit and other cosmological data. 

This paradox could be also resolved by the introduction of a much smaller
Hubble constant, as is advocated by Blanchard et al. \cite{r2} for the Einstein-de Sitter
model. Actually, the estimate of the Hubble constant is sensitive not only
on the cosmological model, but also on our detailed knowledge of the
mechanisms of standard candles, their galactic environment 
and their evolution \cite{r3,r4}.
One can expect substantial changes and improvements in this field of research.

The Einstein-Cartan(EC) gravity, as a quantum theory of gravity, predicts a
negative cosmological constant \cite{r5} assuming the CDM particle to be
a fermion and the scale of weak interactions to be a fundamental scale \cite{r6}.
The EC gravity is the only theory of gravity with definite prediction 
for the cosmological constant.
The phenomena of vorticity and acceleration appears naturaly within the EC
cosmology \cite{r5}. The observed anisotropy in WMAP data has now been
confirmed as real but its cosmological origin needs to be clarified \cite{r7}.

In this note we deal with the scale-noninvariant primordial 
spectrum inferred from WMAP \cite{r1} that has a 
small negative tilt index as a deviation from the 
scale-invariant Harrison-Zeljdovi\v c-Peebles spectrum.
Instead of assuming constraint for the inflationary scenarios at
a scale of $O(10^{-31} cm)$, a scale that is much smaller than
the EC cosmology and weak interaction fundamental scale $O(10^{-16} cm)$, we
assume a scale-invariant spectrum \cite{r8} at a decoupling scale of
$O(10^{25} cm)$ and search for a change of the evolving observables
in the histogram within an inhomogeneous cosmology up to the present $O(10^{28} cm)$.  

To test this idea one could study the CMBR fluctuations beyond
Friedmann geometry using an improved CAMB code based on the Ellis-Bruni
covariant and gauge-invariant approach \cite{r9}.
We perform instead the Press-Schechter(PS) analysis in order to
constrain the acceleration parameter \cite{r10}.



From the Press-Schechter mass functions one can derive 
observables like subhalo mass functions, cumulative
mass functions, galaxy densities, etc \cite{r11}.
We use the Sheth and Tormen(ST) form \cite{r12} 
confirmed by numerical simulations with cold dark matter.
It is necessary to have an evolution equation for
the cosmic mass-density contrast for arbitrary 
inhomogeneous models and then to incorporate the normalized
solutions into the ST functions.

In Ref. \cite{r13} one can find 
the derivation of the evolution for the
density contrast as a function of the cosmic 
mass-density, cosmological constant
and the acceleration parameter $\Sigma$
that describes the inhomogeneity of spacetime:

\begin{equation}
ds^{2} = dt^{2}-a^{2}(t)[(1-\Sigma)dr^{2}+r^{2}(d\theta^{2}
+sin^{2}\theta d\phi^{2})]-2\sqrt{\Sigma}a(t)dr dt, 
\end{equation}
\begin{equation}
a^{\mu}\equiv u^{\nu}\nabla_{\nu} u^{\mu},\ a^{\mu}a_{\mu}
= -\Sigma \frac{\dot{a}^{2}}{a^{2}}, \dot{a}\equiv \frac{da}{dt},
\end{equation}
\begin{eqnarray}
&&\delta^{\prime\prime}+\frac{3}{2}a^{-1}(\Omega_{m}a^{-3}+
2\Omega_{\Lambda})(\Omega_{m}a^{-3}+\Omega_{\Lambda})^{-1}
\delta^{\prime}  \\
&&-\frac{3}{2}a^{-5}\Omega_{m}(\Omega_{m}a^{-3}+\Omega_{\Lambda})^{-1}
\delta -\frac{3}{2}\Omega_{m}^{4/3}\Sigma^{1/2}a^{-6-1/4}
(\Omega_{m}a^{-3}+\Omega_{\Lambda})^{-3/2}\delta=0, \nonumber
\end{eqnarray}
\begin{equation}
\delta^{\prime}\equiv \frac{d \delta}{d a},\ a(z)=1/(1+z).
\end{equation}

In the derivation of 
this equation the term with $\Sigma$
contains R-dependence explicitly and
violates the scale-invariance \cite{r13}.
It is easy to verify that the solution to
this equation is reduced to the standard one
for vanishing $\Sigma$ and positive cosmological
constant:

\begin{equation}
\delta (a)= \frac{1}{x_{0}}\frac{\sqrt{1+x^{3}}}{x^{3/2}}
\int^{x}_{0}\frac{{\rm d}x x^{3/2}}{(1+x^{3})^{3/2}},\  
x=x_{0}a,\ x_{0}=(\frac{\Omega_{\Lambda}}{\Omega_{m}})^{1/3}
\end{equation}

The detailed form of ST mass function that will be used 
in our analysis \cite{r14} is:

\begin{equation} 
n_{ST}(M) dM = A \Big( 1+\frac{1}{\nu^{2q}}\Big) \sqrt{\frac{2}{\pi}} \frac{\rho_m}{M} \frac{d\nu}{dM} {\rm exp}\Big(-\frac{\nu^2}{2}\Big) dM\, ,
\end{equation}

\noindent with $\nu=\sqrt{a_{0}}\frac{\delta_c}{D(a) \sigma(M)}$,
$a_{0}=0.707$, $A\approx 0.322$ and $q=0.3$; $\sigma(M)$ is the
variance on the mass scale $M$, 
$\delta_c$ is the linear threshold for spherical collapse
for a flat universe ($\delta_c=1.686$), 
$\varrho_{\rm m}$ is the background density, 
$D(a)=\delta(a)/\delta(1)$ is the linear growth factor normalized as $D(1)=1$,
$W_{\rm TH}(k R)$ is a top-hat filter in Fourier space with
R defined by $M=4 \pi \varrho_{\rm m} R^{3}/3$ and:

\begin{equation}   
    \sigma^2 = \frac{1}{(2 \pi)^3} \int {\rm d}^3 k P(k)
    W^2_{\rm TH}(k R),
\end{equation}
\begin{equation}
   W_{\rm TH}(x) = \frac{3}{x^{2}} (\frac{\sin x}{x} -\cos x).
\end{equation}

The present-time ($a=1$) power
spectrum is $P(k) = C k^{1+\alpha_{t}} T^2 (k)$ 
where $C$ is the normalization, $\alpha_{t}$ the index of tilt
and $T(k)$ is the transfer function. 
We take the transfer function $T(k)$
in the following form 

\begin{eqnarray}
   T^2(k) &=& \frac{\ln^2(1+2.34 p)}{(2.34 p)^2}  \\
   & \times & [1 + 3.89 p + (16.1 p)^2
   + (5.46 p)^3 + (6.71 p)^4]^{-1/2} \nonumber
\end{eqnarray}
where $p=k/(\Gamma h_{0} {\rm Mpc}^{-1})$, $\Gamma=\Omega_m h_{0} \exp [-\Omega_{\rm b}
(1+ \sqrt{2 h_{0}}/\Omega_m)]$, $\Omega_{b}$ is the baryon density and
$h_{0}$ is the Hubble parameter.

For the spectra normalized to $\sigma_8$ (the
rms density fluctuation smoothed with $R=8 h_{0}^{-1}$ Mpc)
there is no need to introduce
a correction factor (see ref. \cite{r11,r14}) because it does not affect
the calculation of $\sigma$.

Our aim is to find the best fit to the ST mass functions defined by a
tilted primordial spectrum in the Friedmann geometry. Thus, 
we introduce the following deviation measure $\lambda$:

\begin{eqnarray}
\lambda &=& [\frac{1}{i_{max}\times j_{max}}
\sum_{i=1}^{i_{max}}\sum_{j=1}^{j_{max}}\epsilon ^{2}_{ij}]^{1/2},
\\
\epsilon _{ij} &=& [n_{ST}^{(1)}(M_{i},z_{j})-
n_{ST}^{(2)}(M_{i},z_{j})]/n_{ST}^{(1)}(M_{i},z_{j}). \nonumber
\end{eqnarray}

Here $n_{ST}^{(1)}$ represents $ST$ functions with a tilted
spectrum index $\alpha_{t} \neq 0$ and $\Sigma = 0$, 
$n_{ST}^{(2)}$ has $\alpha_{t} = 0$ and $\Sigma \neq 0$, while
the parameters $\Omega_{m},\ \sigma_{8},\ h_{0}$ 
remain unaltered.
The summations are over redshifts discretized linearly and 
over masses logarithmically.
Because of the limited validity of the PS formalism
we use $z_{max}=4$, and owing to the normalization of
the growth function to redshift zero and low sensitivity
at small redshifts,
we arbitrarily set $z_{min}=0.5$.

Using the parameters of WMAP \cite{r1} we plot the
deviation measure $\lambda$ as a function of the acceleration
parameter $\Sigma$ (see Fig. 1). The minimum represents (see Fig. 2) the best fit,
$n_{ST}^{(2)}$, and gives us a clue on how large
the acceleration parameter $\Sigma$ could be 
($\Sigma(\lambda_{min})=0.353\times 10^{-3}$).

Varying upper and lower bounds on the redshift we obtain
(rendering all other parameters as in Fig. 1)
$\Sigma (\lambda_{min})=0.371\times 10^{-3}$ for $z \in [1,4]$
and $\Sigma (\lambda_{min})=0.167\times 10^{-3}$ for $z \in [0.5,3]$.

The dependence of $\Sigma(\lambda_{min})$ on $\sigma_{8}$ is
not significant but it depends sensitively on the 
magnitude of the index of tilt, as an example $\Sigma(\lambda_{min})=1.08\times
 10^{-3}$
for $\alpha_{t}=-0.08$ (other parameters as in Fig. 1).

Similarly, one can find the results for the Einstein-de Sitter and
Einstein-Cartan cosmologies:
$\Omega_{m}=1,\ h_{0}=0.5,\ \Sigma(\lambda_{min})=1.20\times 10^{-3}$ and
$\Omega_{m}=2,\ h_{0}=0.4,\ \Sigma(\lambda_{min})=2.05\times 10^{-3}$
(other parameters as in Fig. 1).  

To conclude, it seems that it is possible to constrain 
the acceleration parameter $\Sigma$ from the WMAP data within
a PS formalism to be of the order of $O(10^{-3})$. This is in
accordance with the estimate of vorticity $\omega_{0}=O(10^{-13} yr^{-1})$ \cite{r15} and
the relationship in the Einstein-Cartan cosmology $\omega_{0} \simeq \Sigma H_{0}$
\cite{r5}.
As mentioned above there is a universal fundamental constant of distance for
both particle physics and EC cosmology.
Therefore, which theory of gravity correctly describes physical reality,
may be determined very soon by the pilot run of the LHC 
 in 2007 \cite{r6,r16}.

\begin{figure}
\begin{center}
\epsfxsize=12cm
\epsfbox{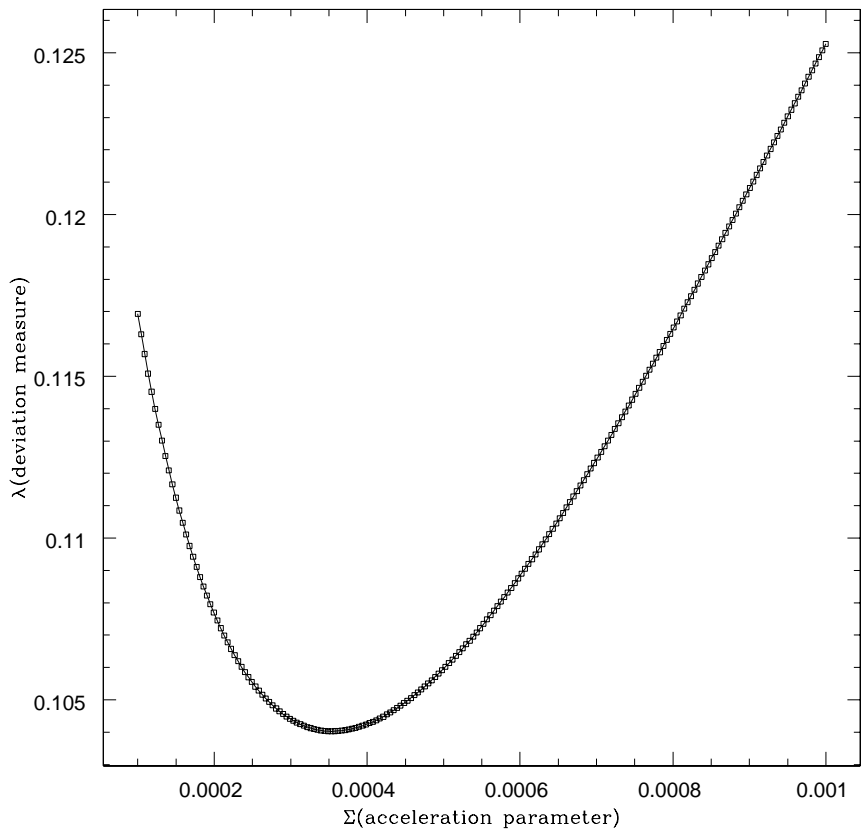}
\caption{Deviation measure ($\lambda$) as a function
of the acceleration parameter ($\Sigma$) for $\Omega_{m}=0.266$,
 $\Omega_{b}=0.044$, $\sigma_{8}=0.77$, $h_{0}=0.71$,
 $\alpha_{t}=-0.05$, $z\in [0.5,4]$,
 $M\in [10^{14},10^{16}]M_{\odot}$, $i_{max}=j_{max}=20$.}
\end{center}
\end{figure}

\begin{figure}
\begin{center}
\epsfxsize=12cm
\epsfbox{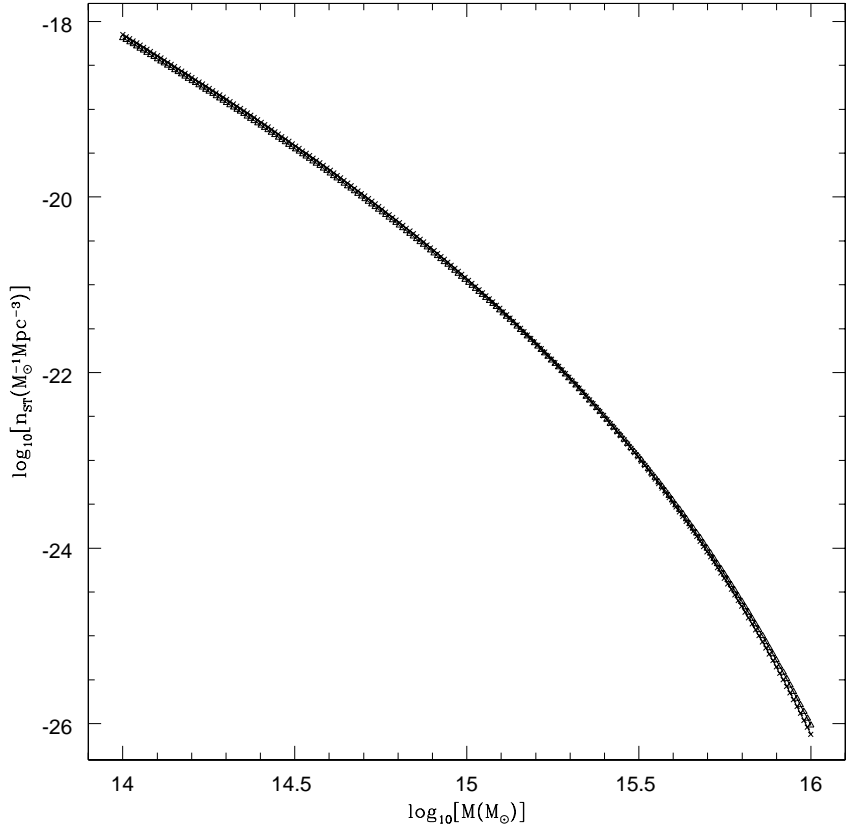}
\caption{ $n_{ST}(M)$ mass functions presented at $z=2$
 for parameters of Fig. 1, where
curve denoted by triangles(crosses) 
is $n_{ST}^{(1)}$($n_{ST}^{(2)}(\Sigma(\lambda_{min}))$).}
\end{center}
\end{figure}

{\sl Acknowledgment}
\\
It is a pleasure to thank Dr. P. Higbie and Prof. A. Klypin for hospitality
at Astronomy Department, NMSU, Las Cruces and fruitful discussions.

\end{document}